\documentclass[tinyml]{acmart}
\AtBeginDocument{%
  \providecommand\BibTeX{{%
    \normalfont B\kern-0.5em{\scshape i\kern-0.25em b}\kern-0.8em\TeX}}}

\setcopyright{rightsretained}
\copyrightyear{2023}
\acmYear{2023}

\begin{document}

\title{Device management and network connectivity as missing elements in TinyML landscape}

\author{Tomasz Szydlo}
%%%\authornote{Both authors contributed equally to this research.}
\email{tomasz.szydlo@{agh.edu.pl, newcastle.ac.uk}}
%%%\orcid{1234-5678-9012}
%%%\author{G.K.M. Tobin}
\authornotemark[1]
%%%\email{webmaster@marysville-ohio.com}

\affiliation{%
  \institution{\textit{Newcastle University, Newcastle upon Tyne, UK}}
  %\streetaddress{Mickiewicza 30}
  %\city{Newcastle Upon Tyne}
  %%%\state{Ohio}
  %\country{Poland}
}

\affiliation{%
  \institution{AGH-University of Science and Technology, \\ Krakow, Poland}
  %\streetaddress{Mickiewicza 30}
  %\city{Cracow}
  %%%\state{Ohio}
  %\country{Poland}
  %\postcode{32-089}
}

\author{Marcin Nagy}
\email{m.nagy@avsystem.com}
\affiliation{%
  \institution{AVSystem}
  \streetaddress{Radzikowskiego 21}
  \city{Krakow}
  \country{Poland}}

%\affiliation{%
%  \institution{Inria Paris-Rocquencourt}

%\affiliation{%
% \institution{Rajiv Gandhi University}

%\affiliation{%
%  \institution{Tsinghua University}
%  \streetaddress{30 Shuangqing Rd}

%\affiliation{%
%  \institution{Palmer Research Laboratories}
%  \streetaddress{8600 Datapoint Drive}
%  \country{USA}

%\affiliation{%
%  \institution{The Th{\o}rv{\"a}ld Group}
%  \streetaddress{1 Th{\o}rv{\"a}ld Circle}
%\email{jsmith@affiliation.org}

%\affiliation{%
%  \institution{The Kumquat Consortium}
%  \city{New York}
%\email{jpkumquat@consortium.net}

\renewcommand{\shortauthors}{Szydlo and Nagy}

\newcommand{\marcin}[1]{\textcolor{blue}{#1}}

\begin{abstract}
Deployment of solutions based on TinyML requires meeting several challenges. These include hardware heterogeneity, microprocessor (MCU) architectures, and resource availability constraints. Another challenge is the variety of operating systems for MCU, limited memory management implementations and limited software interoperability between devices. A number of these challenges are solved by dedicated programming libraries and the ability to compile code for specific devices. Nevertheless, the challenge discussed in the paper is the issue of network connectivity for such solutions. We point out that more emphasis should be placed on standard protocols, interoperability of solutions and security. Finally, the paper discusses how the LwM2M protocol can solve the identified challenges related to network connectivity and interoperability.
\end{abstract}

\keywords{TinyML, LwM2M, IoT interoperability, device management}

\maketitle

\section{Introduction}
The advances in machine learning have a tangible impact on developing the Internet of Things systems. However, the real breakthrough is using ML models directly on microcontrollers, which is possible thanks to TinyML algorithms. It opens new possibilities in various verticals, including industrial predictive maintenance, asset monitoring, street lighting, water and waste management, and transportation. TinyML enhances cloud-based ML by enabling deployments in remote locations and improves user privacy.
%\marcin{Advantage of running ML without cloud dependency: deployments in remote locations, privacy enhancements. Let's not discard standard ML, but emphasize the different type of application}

%However, the benefits of TinyML in smart devices can be enhanced by robust network connectivity. 
However, it is difficult to explore the true potential of TinyML without complementing it with network connectivity mechanisms.
Various energy-efficient communication technologies are designed to cope with the scale of IoT deployments. For example, low-power WAN (LPWAN) solutions enable energy-efficient communication for large geographical areas. Technologies such as NB-IoT and LTE-M are based on a cellular infrastructure with high signal coverage and moderate cost of communication modules. In contrast, LoRaWAN requires dedicated gateways infrastructure, but communication modules are more power efficient and cheaper, though the radio communication spectrum is not licensed. The promising enablers for this kind of IoT devices are SiP (System-in-Package) modules such as nRF9160, STM LBAD0ZZ1SE or STM32WL55 as presented in the Tab.~\ref{tab:devices}. They integrate LPWAN communication modems and embedded ARM cores that simplify the development of smart ML-enabled IoT devices and shorten the time-to-market. In addition to network connectivity, TinyML deployments require as well operational tools that must be adapted to these constraints.

%\marcin{I'd emphasize the need for operational tools for TinyML deployments. Network connectivity is a must for these tools.}

% \usepackage{graphicx}
\begin{table}[]
\centering
\caption{Modules with integrated radio transceivers.}
\label{tab:devices}
\resizebox{\columnwidth}{!}{%
\begin{tabular}{|l|l|l|l|}
\hline
\textbf{MCU/SiP} & \textbf{Clock} & \textbf{Memory}        & \textbf{Radio} \\ \hline
NRF9160          & 64MHz          & 1MB Flash, 256kB RAM   & NB-IoT, LTE-M \\ \hline
NRF52840         & 64MHz          & 1MB Flash, 256kB RAM   & BLE, 802.15.4  \\ \hline
LBAD0ZZ1SE       & 80MHz          & 512kB Flash, 160kB RAM & NB-IoT, LTE-M \\ \hline
STM32WL55        & 48MHz          & 256kB Flash, 64kB RAM  & LoRa           \\ \hline
ESP32            & 240MHz         & 4MB Flash 520kB RAM    & WiFi, BLE      \\ \hline
\end{tabular}%
}
\vspace{-0.4in}
\end{table}

The most important operational tools are related to
%Building robust intelligent IoT solutions based on machine learning on a large scale involves several challenges that must be addressed. 
device management, i.e., service enablement, ML model updates according to TinyMLOps and secure operations.
%and the ones providing a semantic layer for data interoperability between various verticals. 
Furthermore, the size of the TinyML models on devices should be small enough to accommodate the communication stack, device drivers, and the application logic in the device's flash memory, leaving enough space for the buffers required for secure firmware updates. Given the limited resources available on devices, these requirements should be considered holistically, calling for the proper selection of TinyML algorithms and communication protocols. There is also a need for standardized communication stack that can serve multiple services and multiplex connections to conserve device resources. Standardized communication enables building an open ecosystem of IoT solutions and this is what IoT market desperately needs now.

%\marcin{Need for standardized communication stack that can serve multiple services and multiplex connections to conserve resources. Standardized communication enables building an open ecosystem of IoT solutions and this is what IoT market desperately needs now.}

%\marcin{In this paper, we explore technical mechanisms}
In the paper, we explore technical mechanisms for providing operational support of TinyML deployments. We focus on the discussion of real-world requirements and also also present an implementation of a real-world solution.

%\marcin{Outline. What do we demonstrate in this work}
The paper is organized as follows. In the Section 2 related work is presented. Then in the Section 3, we discuss the requirements for a succesful deployment of TinyML. Section 4 discusses TinyML lifecycle management, while sections 5 and 6 discuss the applicability of LwM2M protocol. Section 7 discusses the case study. Finally, section 8 concludes the paper.

\section{Related work}
There is a growing interest in the field of TinyML, which involves running machine learning models on devices with limited resources, such as single-board computers, including Raspberry Pi and Nvidia Jetson Nano and the smaller ones based on microcontrollers. On devices with microprocessors running Linux, various techniques are used to reduce the size and processing time of neural networks, such as quantizing weights \cite{quantization}, training the binary networks \cite{binary}, or using NN accelerators. However, an even more challenging research problem is applying these techniques to millions of Internet of Things (IoT) devices equipped with microcontrollers (MCUs) that have limited resources such as 1 MB of flash memory and 256 KB of RAM. To address this issue, TinyML solutions may involve using different neural network architectures e.g. Once-For-All network \cite{ofa}, preprocessing digital signals, and exploring the use of traditional machine learning methods such as decision forests.

Secondly, obtaining data from devices in real-world environments involves communication technologies and protocols. The 5G network provides various communication profiles for applications, such as low latency or high volume communication. However, effectively using these profiles requires research in network function virtualization (NFV) and edge computing \cite{5G}. Low power wide area networks (LPWANs) like NB-IoT, LTE-M, and LoRaWAN are equally important for gathering data from IoT devices. Research on optimizing communication protocols is crucial to minimize energy use and adapt to networks' conditions. In addition, developing mesh networks like Thread\footnote{https://www.threadgroup.org/} and Matter, which rely on standards like 802.15.4 for communication between devices, requires research on network topology and cost-effective binary communication protocols like LwM2M, CoAP, and CBOR.

Finally, some enterprises such as Queexo\footnote{https://qeexo.com/}, Edge Impulse\footnote{https://www.edgeimpulse.com/}, SensiML\footnote{https://sensiml.com/} and others are democratising TinyML by providing developers with easy-to-use tools. However, they focus primarily on delivering AutoML rather than interoperable and network-connected solutions.

\section{System requirements}
Our view on a successful TinyML solution sets four basic goals to fulfil:
\begin{enumerate}
    \item Usability $\Rightarrow$ the system should be easy to operate, thus necessary ops tools need to included in the design.
    \item Interoperability $\Rightarrow$ the system should enable to be integrated with various IoT solution providers.
    \item Sustainability $\Rightarrow$ the system should efficiently manage resources to maximize its lifetime.
    \item Security $\Rightarrow$ assurance of system security is a critical requirement to convince users to TinyML systems.
\end{enumerate}

Based on that, we set the following requirements the IoT system should fulfil:

\textbf{Device firmware update.} 
% removing bugs
% updating ML for better accuracy
% local optimizations of ML models
% efficient processes
Since bugs are ubiquitous, device firmware must be upgraded once in a while. Device software containing optimized neural networks needs to be regularly updated to improve its accuracy. Finally, numerous environmental factors may require local optimizations of neural networks. Such factors are often discovered only once a device is deployed in the field.    

\textbf{Data interoperability.}
IoT is a very fragmented market with many solution suppliers and consumers. To provide a better value to all these entities, build scalable solutions and create synergy between them, data must be well understood by all parties. This motivates building interoperable solutions.

\textbf{Efficient cloud communication.}
Device cost comprises two parts: 1) a one-time manufacturing cost and 2) a recurring operational cost. The manufacturing cost is proportional to the amount of physical resources a device has, especially the flash, memory and battery size. The operational costs include firmware updates, connectivity and data collection costs. The former can be optimized by the usage of resource-efficient network libraries. The latter can be optimized by using communication protocols for constrained communication environments, for instance, Low-Power WAN (LPWAN). Finally, it is also important to note that efficient cloud communication positively impacts device battery lifetime.   

\textbf{System security.}
System security consists of two requirements: 1) cloud communication channel protection and 2) secure execution of on-device operations. To be trusted, they should use well-established technologies and software frameworks rather than vendor-specific solutions.

\section{LwM2M protocol} %\marcin{Maybe move it before TinyML Lifecycle management?}
We argue that the aforementioned requirements related to device management and communication can be handled by a standardised LwM2M protocol\footnote{https://technical.openmobilealliance.org/OMNA/LwM2M/LwM2MRegistry.html} designed to manage millions of IoT devices leveraging LPWAN technologies. Moreover, it has several mechanisms enabling the implementation of intelligent IoT solutions on a global scale, such as object data model, firmware updates and secure communication.

The LwM2M is an application-level protocol for device management and service enablement. Due to resource limitations of IoT devices, it was built based on the Constraint Application Protocol (CoAP). As a result, it can be used e.g. in low-power mesh networks such as Thread or Matter. LwM2M also supports various binary data serialisation mechanisms, enabling data transmission in human-readable JSON and a binary form based on CBOR.

LwM2M also supports a communication mode called queue mode, which is useful when the device is not always available via the LwM2M server due to long periods of sleep to preserve energy in the device batteries. These mechanisms are, to some extent, transparent to the end-user, simplifying system design.

%communication
%serialization
%notification
%semantics
The requirement for device firmware update is fulfilled by the LwM2M protocol - specification introduces the /ID5 object responsible for updating the device firmware. Two modes of operation are identified - \textit{Push} and \textit{Pull} during which various communication protocols can be used for transferring firmware i.e. CoAP/UDP, CoAP/TCP, and HTTP, depending on the application.
%The LwM2M specification introduces the /ID5 object responsible for updating the device firmware. 
%\marcin{I'd remove part on push and pull FOTA. I'm not sure that it adds much value to the text.}
%The specification defines two modes of firmware download: 
%\begin{itemize}
%    \item PUSH - the download is initiated by the Server, performing a write operation on the resources of ID/5 object exposed by the device. This usually means a block-wise transfer provided by the CoAP protocol.
%    \item PULL - the LwM2M Server indicates the URL from which the device should download the firmware package. It is done by performing write operation on resources of object ID/5. Then the client then performs the download asynchronously. In this case, several protocols can be used for downloading the firmware i.e. CoAP/UDP, CoAP/TCP, and HTTP.
%\end{itemize}

The case study shows how it can be used to update an IoT device. The method of reprogramming the device and changing the firmware version depends on, e.g. on the hardware, bootloaders, and memory organization. At the level of the semantics of LwM2M objects, the /ID5 object is responsible for sending a new version of the firmware to the device.

From the point of view of managing TinyML on devices, information related to updating ML models, reporting the results of their operation and assessing the quality of the results provided by ML models is crucial. The last two requirements are discussed in more detail in the next section.

\section{TinyML lifecycle Management}
LwM2M data model represents, among others, device status, software version, sensors installed in the device, and telemetry data. At the same time, it ensures interoperability between device manufacturers and IoT system providers. As TinyML algorithms enable continuous sensor data processing, it is possible to detect complex patterns and anomalies. However, from the end-user point of view, reported information that a particular event has occurred is usually more important than the algorithm used for its detection, which may change with the new firmware update due to new ML methods development. Therefore, the data model of the LwM2M protocol can be extended with the objects used for the management of TinyML models and supporting the TinyMLOps process \cite{tinymlops}. For this purpose, such objects as \textit{PatternDetector}, \textit{AnomalyDetector}, \textit{Classifier} and \textit{ML model} might be provided and standardised - they are discussed in the next sections.

Using machine learning algorithms on devices and delivering high-quality solutions requires device management and service enablement on a scale that has never been done before. Device firmware with optimized machine learning models should be updated to keep the models up-to-date. It can be achieved by partial or full reprogramming of the device's internal memory. Therefore, firmware update mechanisms, as provided by LwM2M, are essential to ensure the continuity of device operation by performing recovery in unforeseen situations and carrying out firmware rollout efficiently.

\begin{figure}[h]
  \centering
  \includegraphics[width=\linewidth]{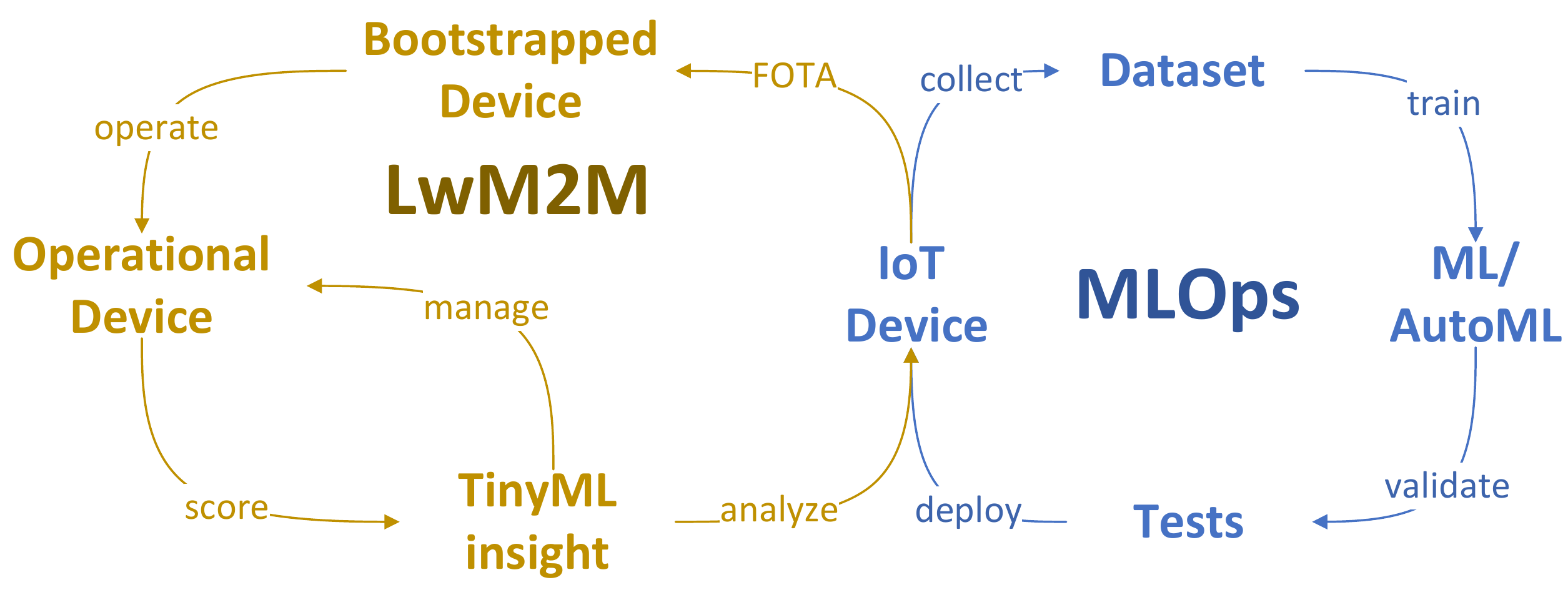}
  \caption{IoT device with TinyML management process.}
  \label{fig:mlops}
  \vspace{-0.1in}
\end{figure}

The Fig.~\ref{fig:mlops} shows how the LwM2M protocol enhances the ML model development process for resource-constrained devices. It consists of 4 states:
\begin{itemize}
    \item \textit{IoT device} - A TinyML model is developed and runs on the IoT device. Testing and verification of operations take place mainly in a controlled environment. The device software containing the new version of the ML model can be remotely uploaded to the device via the firmware update mechanism.
    \item \textit{Bootstrapped device} - LwM2M device connects to LwM2M Bootstrap Server during the first or every bootup procedure to initialize the data model and to obtain addresses of LwM2M servers that the device should connect to. LwM2M protocol provides mechanisms for secure end-to-end communication.
    %the LwM2M client ensures the communication of the IoT device and the transmission of the inference results of the TinyML model deployed on the device. LwM2M protocol provides mechanisms for secure end-to-end communication. 
    \item \textit{Operational device} - LwM2M protocol provides device state update to the LwM2M server. Communication mechanisms include several solutions to reduce resource consumption and monthly dataplan. In addition, the results of TinyML operation on the device are constantly updated on the server and can be accessed via a dedicated API by third-party applications.
    \item \textit{TinyML insight} - dedicated LwM2M objects allow to assess of the TinyML model's operating status on the device and, if necessary, to correct operating parameters or for further analysis and development of a new version of the model.
\end{itemize}

\section{LwM2M objects for TinyML} 
%\marcin{Maybe System design}
In the LwM2M protocol, device resources are represented as objects. To connect LwM2M with TinyML, we have developed the dedicated objects that are related to the machine learning algorithms deployed on the devices. Their goal is to shorten the development time for an IoT solution. They are discussed in more detail in the following sections.\footnote{The objects shown are in an object pool reserved for AVSystem}

\subsection{ML Model object}
\begin{figure}[h]
  \centering
  \includegraphics[width=\linewidth]{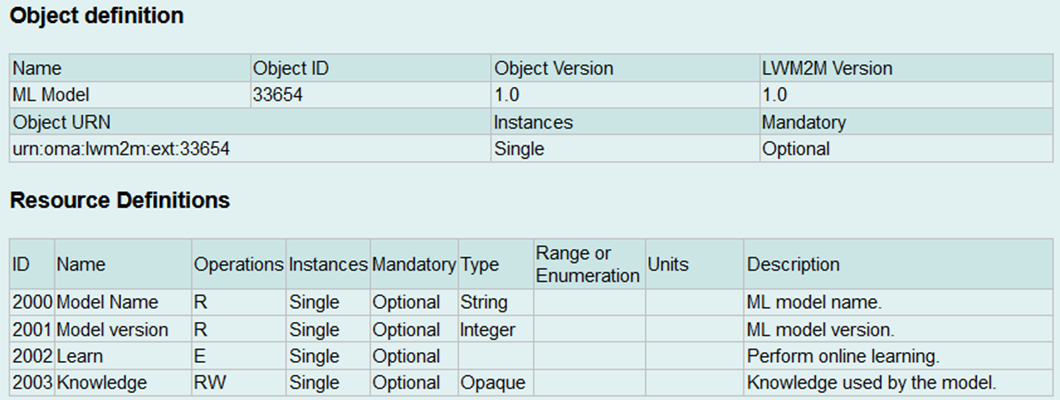}
  \caption{LwM2M object 33654.}
  \label{fig:33654}
  \vspace{-0.1in}
\end{figure}

Object with /ID33654, as presented in Fig. \ref{fig:33654} is used to describe the ML model used on the device. The description consists of the model name and its version. Object /ID33654 have only a single instance, but in the case of complex devices, it can be extended to allow multiple instances. 

\subsection{Pattern Detector object}
\begin{figure}[h]
  \centering
  \includegraphics[width=\linewidth]{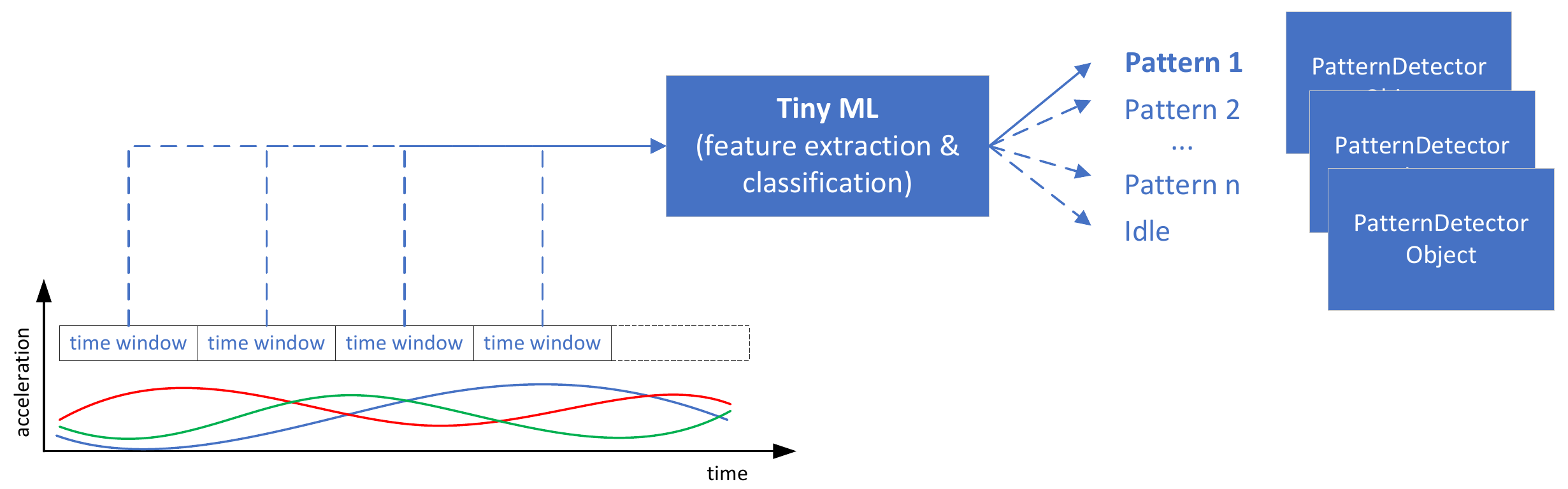}
  \caption{LwM2M object 33650.}
  \label{fig:33650}
  \vspace{-0.1in}
\end{figure}

The object /ID33650, as presented in Fig.~\ref{fig:33650}, represents the results of the pattern classifier applied to the sensor data. An object can have many instances - one for each classifier's output class. It was assumed that for each time window in which sensor data is analyzed, the result of the classification is the class with the highest probability.

\subsection{Anomaly Detector object}
\begin{figure}[h]
  \centering
  \includegraphics[width=\linewidth]{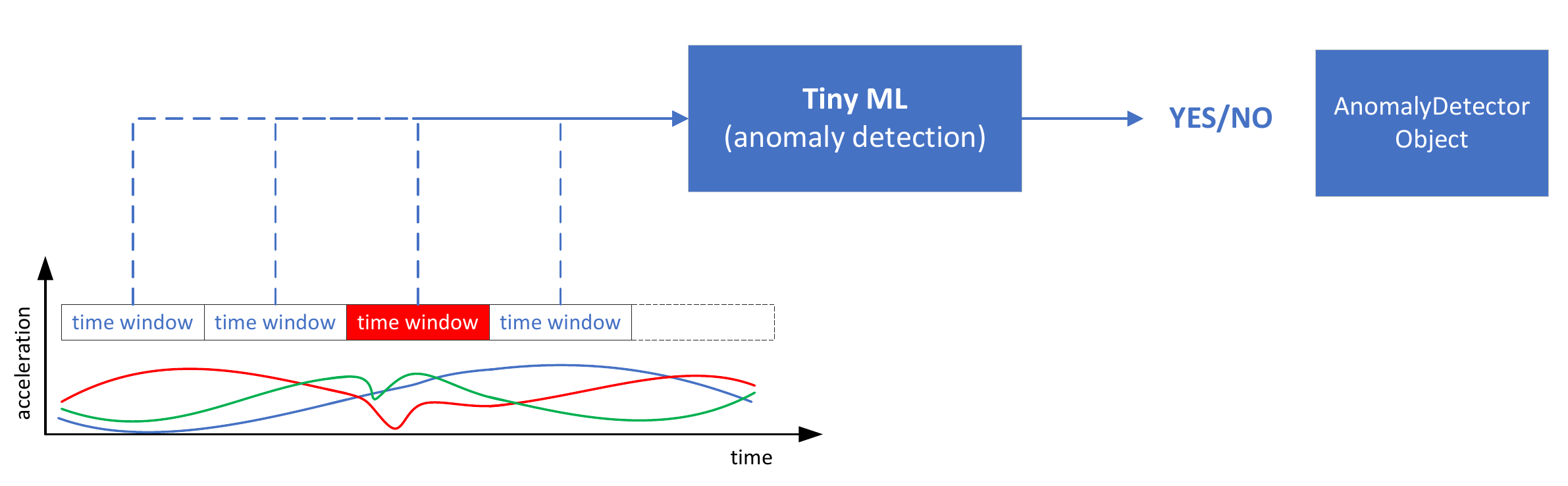}
  \caption{LwM2M object 33651.}
  \label{fig:33651}
  \vspace{-0.1in}
\end{figure}

Object /ID33651, as presented in Fig.~\ref{fig:33651}, is used to report the situations for which the anomaly detection module detected abnormal behaviour. 

The high amount of anomalies found in the sensor data means that the nature of the data is significantly different from the training data. Thus, if the classifier is also running on the device, the classification quality may deviate from the expected values. Many detected anomalies may also indicate that the device works differently from its typical application, e.g. is wrongly used or begins to break down.

\subsection{Classifier object}
\begin{figure}[h]
  \centering
  \includegraphics[width=\linewidth]{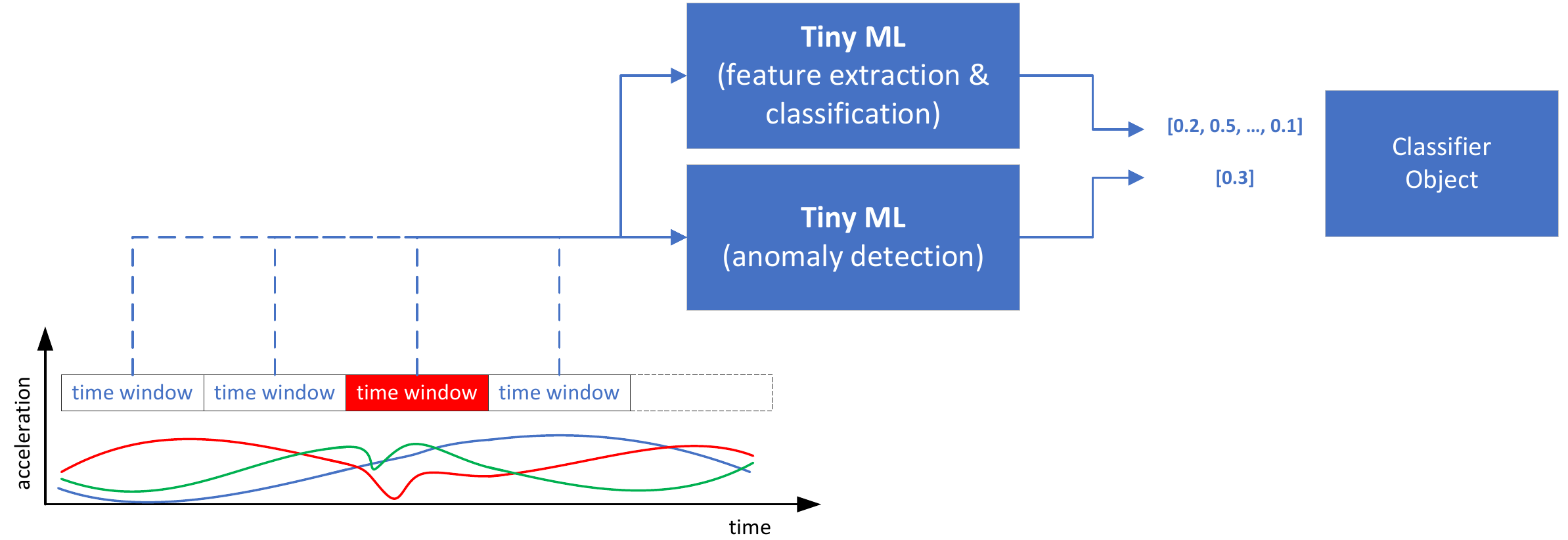}
  \caption{LwM2M object 33652.}
  \label{fig:33652}
  \vspace{-0.1in}
\end{figure}

The object /ID33652, as presented in Fig.~\ref{fig:33652}, is used to report the detailed results of an anomaly detector and a classifier output. The object is updated with each iteration of the ML model operation, which means that the continuous update of the object state on the LwM2M server involves a significant amount of data transferred from the device. On the other hand, it provides the most detailed information of the TinyML outputs on the device.

\subsection{Anomaly Analyzer object}
\begin{figure}[h]
  \centering
  \includegraphics[width=\linewidth]{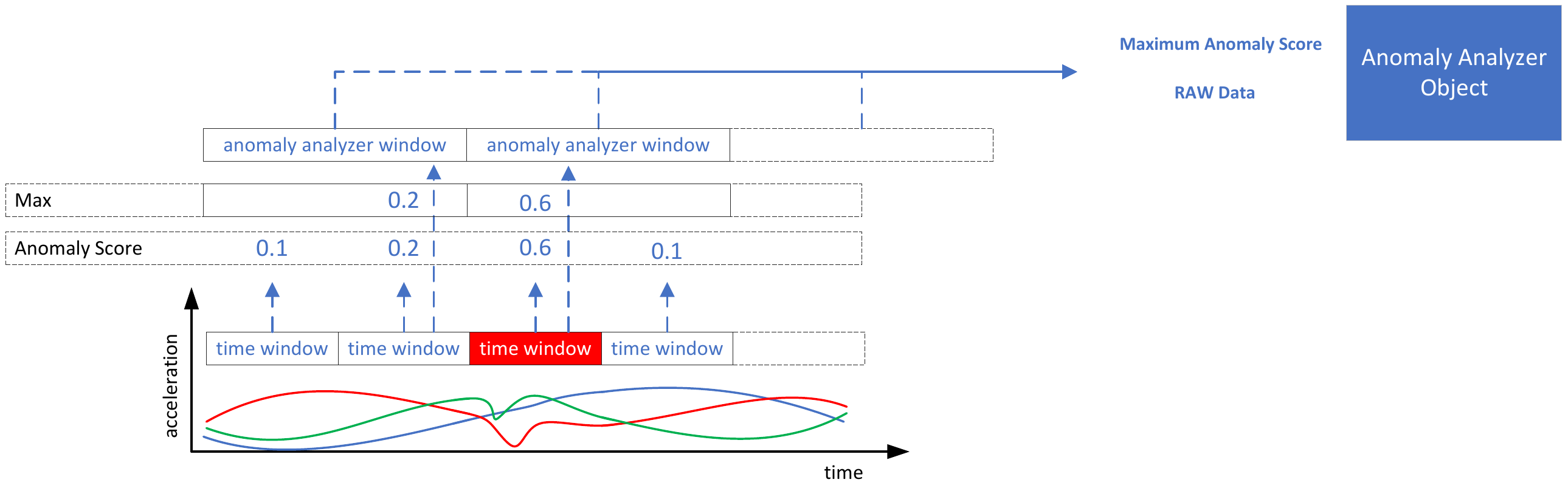}
  \caption{LwM2M object 33653.}
  \label{fig:33653}
  \vspace{-0.1in}
\end{figure}

Object /ID33653, as presented in Fig.~\ref{fig:33653}, is mainly for diagnostic purposes and can be used for TinyML model insight. It aims to indicate specific situations in the device's operation that were not considered in the ML model training process during its development. 

The object should not be used to collect training data for ML models but only indicate areas for deeper analysis during the development of updated versions of the ML model. The object is mainly based on anomaly detection algorithms. In time windows lasting e.g. the whole day, the anomaly score is analyzed, and the sensor data for which the score is the highest is stored. 

An LwM2M object state can be updated regularly, e.g. once a day. As a result, the device will not continuously stream sensor data but only some characteristic data that has occurred during the operation of the device. 

\section{Case Study} %\marcin{What about TinyML solution development: case study}
\begin{figure*}[h]
  \centering
  \includegraphics[width=0.9\linewidth]{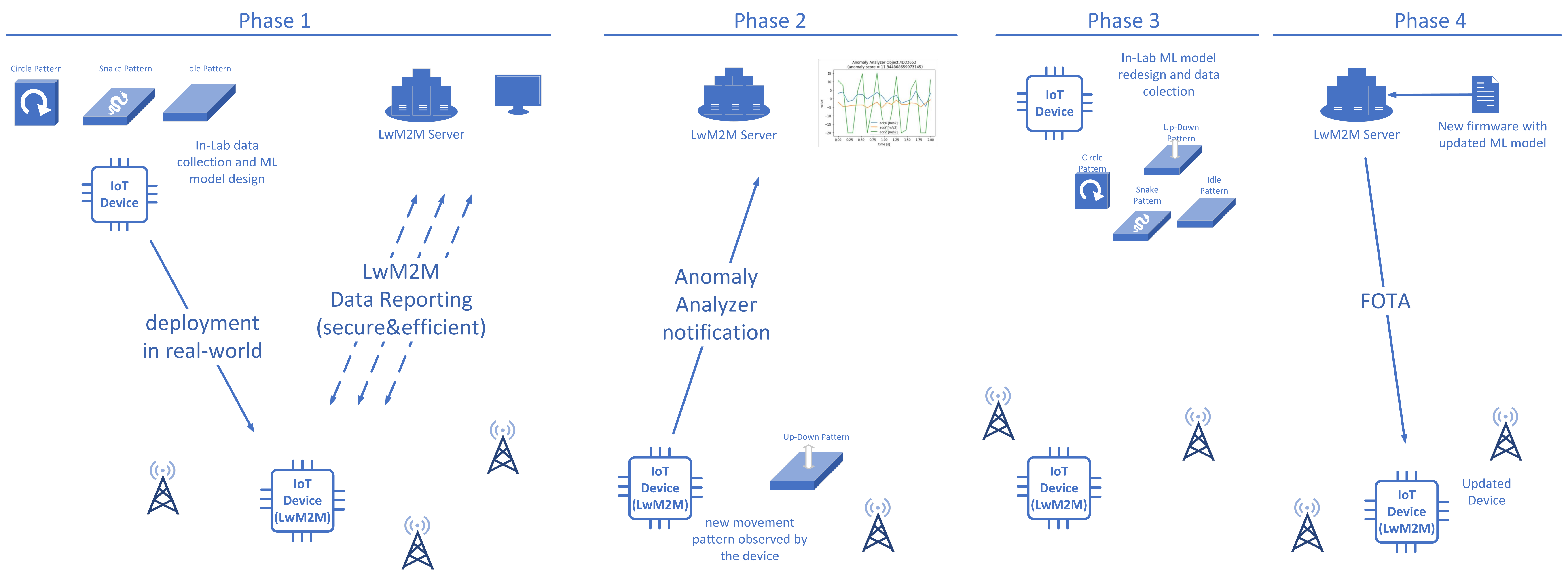}
  \caption{Case study phases.}
  \label{fig:case}
\end{figure*}

The typical development of the application with TinyML, as presented in Fig.~\ref{fig:case}, is divided into phases during which the usage of LwM2M objects are analyzed. The case study is developed for the \textit{Thingy:91} device and uses acceleration sensors as a time-series data source. The functionality of the application includes the detection of characteristic movement patterns. Thanks to the LwM2M Anjay client, the device registers to the LwM2M server, where it updates its state and the inference results of the TinyML model. The source-code of the application is available on Github.\footnote{https://github.com/tszydlo/FogML-Zephyr-FOTA}

\begin{figure}[h]
  \centering
  \includegraphics[width=0.7\linewidth]{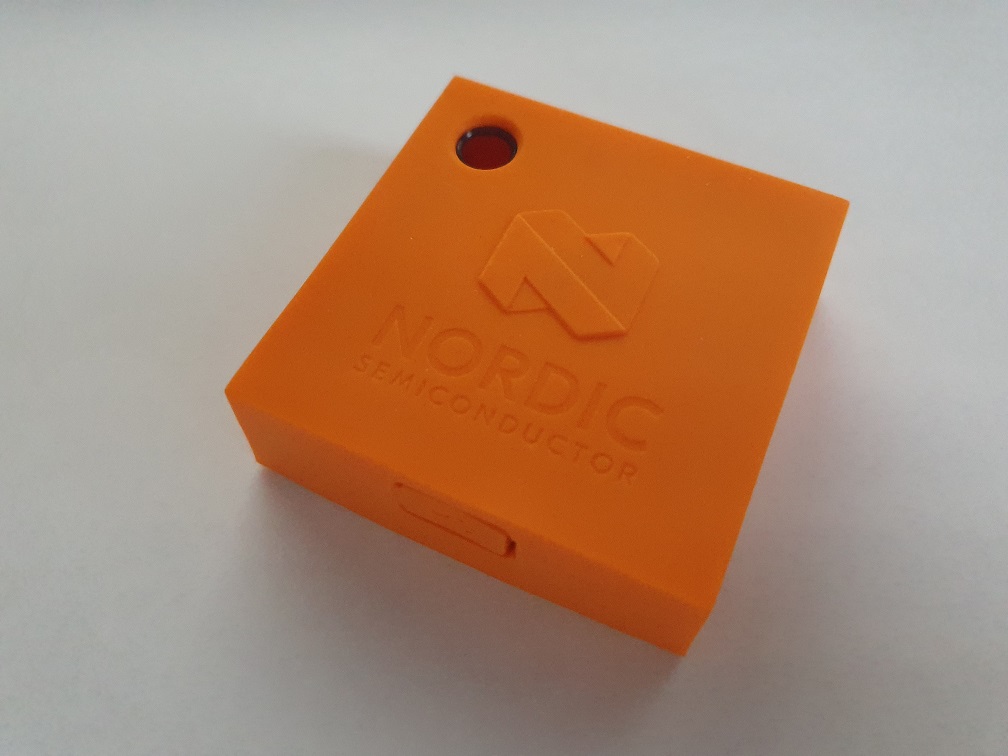}
  \caption{Nordic Thingy:91.}
  \label{fig:thingy}
  \vspace{-0.1in}
\end{figure}

\begin{figure}[h]
  \centering
  \includegraphics[width=\linewidth]{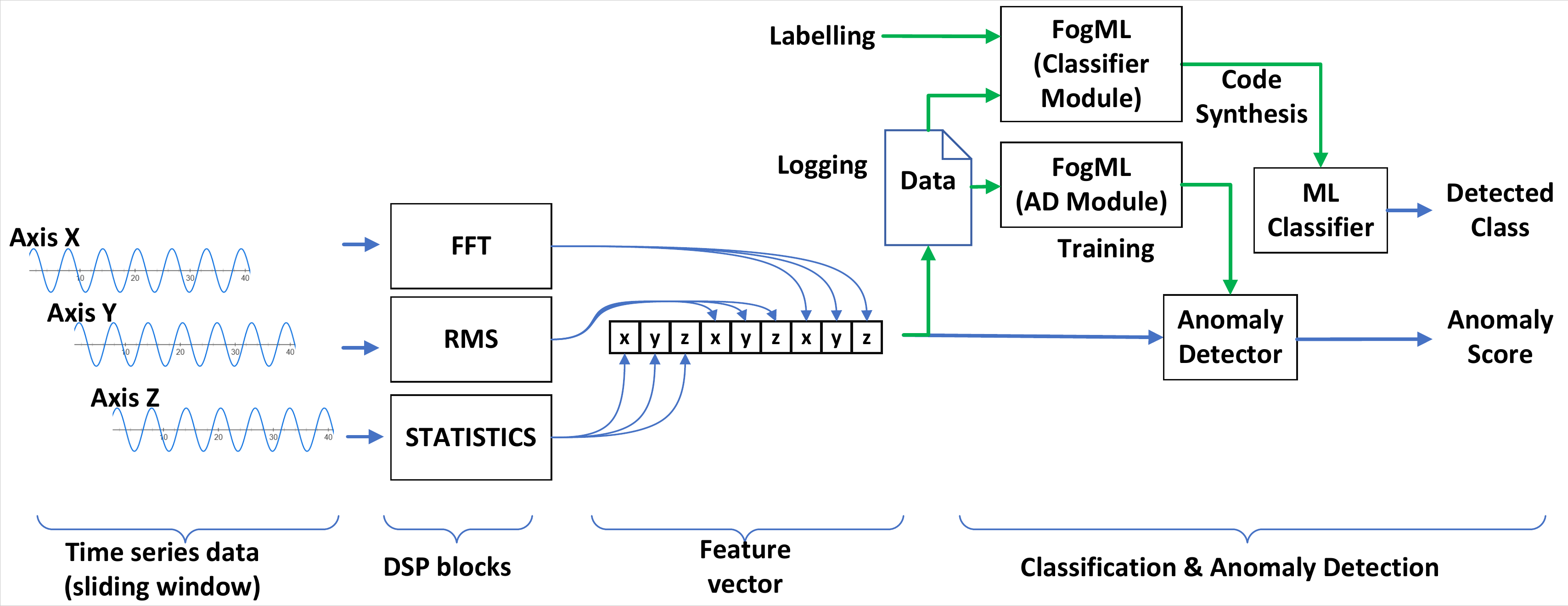}
  \caption{FogML processing used in the case study.}
  \label{fig:processing}
  \vspace{-0.1in}
\end{figure}

\subsection{Thingy:91 device}
Nordic Semiconductor's \textit{Thingy:91} is a cellular-enabled IoT sensor prototyping platform based on the nRF9160 SiP and nRF52840 SoC. It is equipped with several sensors, including accelerometers.

In the example, we only use the NRF9160 SiP, the specification of which is presented in Tab.~\ref{tab:devices}. In the example, we use NRF Connect SDK version 2.0.0, LwM2M Anjay client version 3.0 and FogML tools version 0.7.

The software of the device is based on the Zephyr OS system, additionally extended by Nordic Semiconductor. The device uses MCUboot, a secure bootloader for a 32-bit MCU. Due to the fact that the device does not have an external memory chip, its internal memory is divided into two slots - the first from which the application is launched and the second, which is used to temporarily store the new firmware version.

\subsection{FogML}
\begin{figure}[h]
  \centering
  \includegraphics[width=\linewidth]{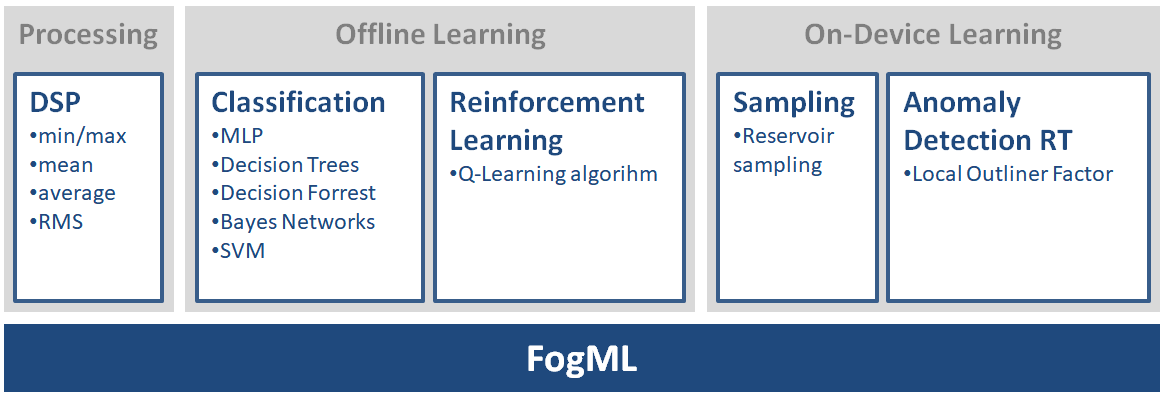}
  \caption{Algorithms supported by FogML.}
  \label{fig:fogml}
  \vspace{-0.1in}
\end{figure}

The FogML (Fig.~\ref{fig:fogml})\footnote{https://github.com/tszydlo/FogML} is a set of tools enabling TinyML on microcontrollers as low resource-limited as ARM M0 cores. In contrast to many other frameworks, the FogML utilises classic machine learning methods such as density-based anomaly detection and classifiers based on Bayesian networks, decision forests, vanilla MLP \cite{fogml_ts}, and reinforcement learning (RL) \cite{TinyRL_ts}. It supports off-device learning for the classification problem and on-device learning for anomaly detection. Active learning anomaly detection is based on reservoir sampling and outlier detection algorithms which are trained directly on the device \cite{online_anomaly_ts}. The dedicated library performs the time series processing on the devices, computing the feature vector consisting of RMS, FFT, amplitude and other low-level signal metrics. One of the techniques used in FogML is source code generation of the inferencing functions for embedded devices. It leads to a much smaller memory footprint than for more computationally advanced solutions such as deep neural networks.

\subsection{Anjay LwM2M client}
%Tools
%custom features
%supported boards and hardware
%linki do repo
%supported networks/NIDD
Anjay is an advanced and specification-compliant LwM2M client. The implementation supports a number of operating systems such as mbed, FreeRTOS, Zephyr OS, esp-idf and RPi Pico. In addition, the client is provided with tools supporting implementation, particularly for creating stubs for LwM2M objects based on their XML specification.

The example of the TinyML application presented in the case study is based on the demo application for the Anjay LwM2M client\footnote{https://github.com/AVSystem/Anjay-zephyr-client}, which has been extended with FogML.

\subsection{Phase 1}
It was assumed that the device should detect three movement patterns, as shown in Fig.~\ref{fig:case}. Using tools provided by FogML, we collected and labelled accelerometer data and then trained the random forest ML model (MODEL1) to detect movement patterns. We have also trained the \textit{k-means} based anomaly detector on the non-labelled training data. This will be used for the TinyML insight - to asses how much real deployment differs from the lab environment. Details of the processing are presented in Fig.~\ref{fig:processing} and Tab.~\ref{tab:models}.

After flashing and restarting, the device establishes the connection to the LwM2M server via NB-IoT. Information about the ML model used on the device is available through the object /ID33654. The detected movement patterns and the counter value of their occurrences are represented as consecutive instances of the object /ID33650.

Usually, the device works in a real environment and can operate in situations not foreseen during its development. Therefore, the LwM2M \textit{Anomaly Detector} object can be used as an indicator of such situations. Furthermore, the threshold value for the anomaly detector object can be determined based on the analysis of the LwM2M \textit{Classifier} object.

% \usepackage{graphicx}
\begin{table}[]
\centering
\caption{Application and ML model details.}
\label{tab:models}
\vspace{-0.1in}
\resizebox{\columnwidth}{!}{%
\begin{tabular}{|l|cc|}
\hline
\textbf{Parameter}            & \multicolumn{1}{c|}{\textbf{MODEL 1}}  & \textbf{MODEL 2}  \\ \hline
Firmware Size [B]             & \multicolumn{1}{c|}{396984}            & 399554            \\ \hline
NCS+LwM2M Client+IP Stack [B] & \multicolumn{1}{c|}{278332}            & 278332            \\ \hline
Application [B]               & \multicolumn{1}{c|}{104517}            & 105807            \\ \hline
FogML SDK [B]                 & \multicolumn{1}{c|}{1669}              & 1669              \\ \hline
FogML Random Forrest [B]      & \multicolumn{1}{c|}{11294}             & 12574             \\ \hline
FogML K-Means Anomaly [B]     & \multicolumn{1}{c|}{848}               & 848               \\ \hline
FogML Scaller [B]             & \multicolumn{1}{c|}{108}               & 108               \\ \hline
FogML Config [B]              & \multicolumn{1}{c|}{216}               & 216               \\ \hline
Number   of detected classes  & \multicolumn{1}{c|}{3}                 & 4                 \\ \hline
ML Model                      & \multicolumn{2}{c|}{Random Forrest   (trees=100, depth=3)} \\ \hline
DSP Blocks                    & \multicolumn{2}{c|}{Base, RMS, Crossings}                  \\ \hline
Sensor acquisition [Hz]       & \multicolumn{2}{c|}{10}                                    \\ \hline
k-means clusters              & \multicolumn{2}{c|}{16}                                    \\ \hline
DSP vector size               & \multicolumn{2}{c|}{12}                                    \\ \hline
\end{tabular}%
}
\vspace{-0.1in}
\end{table}

\subsection{Phase 2}
After observing that the device correctly detects the movement patterns, the device should be faced new and unforeseen movement - in our case, moved up and down several times. Such a pattern has not been previously observed in the training data. 

The increase in the value of the \textit{Anomaly detector} /ID33651 indicates that the sensor data input from the TinyML model is significantly different from the training data, and a deeper analysis is needed. The details of the analysis are presented in the next phase.

\subsection{Phase 3}
\begin{figure}[h]
  \centering
  \includegraphics[width=0.8\linewidth]{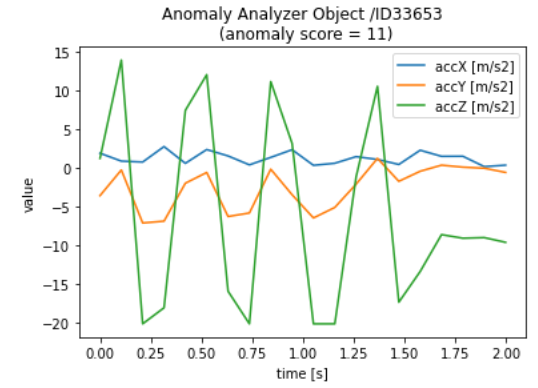}
  \caption{FogML processing used in the case study.}
  \label{fig:anomaly_vis}
  \vspace{-0.1in}
\end{figure}

The object \textit{Anomaly analyzer} /ID33653 enables the access to the sensor data with the highest anomaly score value. Then, data can be visualised using simple Python based tools as presented in Fig~\ref{fig:anomaly_vis}. The values presented in the chart may suggest that the high value of the anomaly score index was achieved for the movement where the acceleration value on the Z-axis changed rapidly.

\subsection{Phase 4}
Analysis of the values represented by the \textit{Anomaly analyzer} object /ID33653 led to the conclusion that the device experienced vertical up-down movement. Therefore, additional training data was prepared in the laboratory, and a movement class was added to the training datasets to represent the identified vertical movement.

The final TinyML model (MODEL2) is compiled with the application, and the new firmware version containing the updated model can be remotely transferred to the devices using FOTA. After the firmware upgrade, the device restarts establishes the connection to the LwM2M server. The process can be repeated to refine the TinyML model and adjust to changing its operational context.

\section{Summary}
To summarise, TinyML solutions are a milestone in developing Internet of Things systems. However, it is necessary to ensure service enablement mechanisms, over-the-air firmware updates, and secure connectivity to cloud computing services to take full advantage of this class of solutions. Furthermore, using low memory footprint ML libraries such as FogML leads to flexibility in designing more complex and connected embedded applications. Finally, requirements related to device management and data reporting can be satisfied by the LwM2M protocol, which solves many technological challenges encountered when implementing IoT solutions on a large scale, especially when connected using LPWAN technologies.

\section{Acknowledgments}
\textit{We would like to thank the embedded team (Mieszko Mieruński, Mateusz Kwiatkowski and Aleksander Wojtowicz) from AVSystem for discussions and technical support regarding the LwM2M Anjay client.}

%This section has a special environment:
%\begin{verbatim}
%  \begin{acks}
%  ...
%  \end{acks}
%\end{verbatim}
%so that the information contained therein can be more easily collected
%during the article metadata extraction phase, and to ensure
%consistency in the spelling of the section heading.
%Authors should not prepare this section as a numbered or unnumbered {\verb|\section|}; please use the ``{\verb|acks|}'' environment.

%\section{Appendices}
%If your work needs an appendix, add it before the
%``\verb|\end{document}|'' command at the conclusion of your source
%document.
%Start the appendix with the ``\verb|appendix|'' command:
%\begin{verbatim}
%  \appendix
%\end{verbatim}
%and note that in the appendix, sections are lettered, not
%numbered. This document has two appendices, demonstrating the section
%and subsection identification method.

\bibliography{acmart}
\bibliographystyle{plainnat}

\end{document}